# DeepSecure: Scalable Provably-Secure Deep Learning

Bita Darvish Rouhani, M. Sadegh Riazi, Farinaz Koushanfar
University of California San Diego
{bita, mriazi, farinaz}@ucsd.edu

## ABSTRACT

This paper proposes DeepSecure, a novel framework that enables scalable execution of the state-of-the-art Deep Learning (DL) models in a privacy-preserving setting. DeepSecure targets scenarios in which neither of the involved parties including the cloud servers that hold the DL model parameters or the delegating clients who own the data is willing to reveal their information. Our framework is the first to empower *accurate* and *scalable* DL analysis of data generated by distributed clients without sacrificing the security to maintain efficiency. The secure DL computation in DeepSecure is performed using *Yao's Garbled Circuit* (GC) protocol. We devise GC-optimized realization of various components used in DL. Our optimized implementation achieves more than *58-fold* higher throughput per sample compared with the best prior solution. In addition to our optimized GC realization, we introduce a set of novel low-overhead pre-processing techniques which further reduce the GC overall runtime in the context of deep learning. Extensive evaluations of various DL applications demonstrate up to *two orders-of-magnitude* additional runtime improvement achieved as a result of our pre-processing methodology. We also provide mechanisms to securely delegate GC computations to a third party in constrained embedded settings.

## KEYWORDS

Deep Learning, Secure Function Evaluation, Garbled Circuit, Content-Aware Data Pre-processing.

## 1 INTRODUCTION

Deep Learning (DL) has provided a paradigm shift in our ability to comprehend raw data by showing superb inference accuracy resembling the learning capability of human brain [1, 2]. Technology leaders such as Microsoft, Google, IBM, and Facebook are devoting millions of dollars to devise accurate DL models for various artificial intelligence and data inference applications ranging from social networks and transportations to environmental monitoring and health care [3–5]. The applicability of DL models, however, is hindered in settings where the risk of data leakage raises serious privacy concerns. Examples of such applications include scenarios where clients hold sensitive private information, e.g., medical records, financial data, or location.

Developing a DL model for a particular task involves two main steps: training and execution. During DL training, the focus is to fine-tune the DL parameters such that the accuracy of the model is maximized. Whereas, the DL execution attempts to find the corresponding inference label for newly arrived data samples using the trained model. To address DL problem in sensitive scenarios, authors in [6, 7] suggest the use of differential privacy for training DL models. As such, training of a DL model is performed by creating a *statistical database* from raw data samples such that the amount of information leaked from an individual record is minimized. The notion of differential privacy, however, is not applicable to the DL execution. This is due to the fact that during the DL execution, one is interested in finding the pertinent output corresponding to *a single data sample* as opposed to the statistical property of a collection of data. In this paper, we focus on devising an end-to-end framework that enables using the trained DL models to analyze sensitive data, while ensuring the confidentiality of both data and DL parameters.

To address DL execution in sensitive applications, Microsoft has recently proposed a cryptographic DL framework (called CryptoNet [8]) based on Homomorphic Encryption (HE). Although this approach enables DL execution in a privacy-preserving setting, it encounters four main limitations that we simultaneously address in this paper: (i) CryptoNet has a privacy/utility trade-off; in order to achieve a higher level of privacy, the utility decreases significantly. In addition, the noise introduced to the DL model as a result of securely encrypting data samples can further lessen DL inference accuracy and yield an incorrect result for the input data that could have been classified correctly otherwise. (ii) Non-linear activation functions which are the key elements in a DL network (e.g., Tangent-Hyperbolic, Sigmoid, etc.) cannot be effectively computed by low-degree polynomials as suggested in [8] using HE. (iii) Homomorphic encryption results in a relatively high computational overhead which bounds CryptoNet's practicability in resource-limited settings where the data owners have severe computational constraints (e.g., smartphones and wearable devices). (iv) CryptoNet incurs a constant computational cost up to a certain number of samples depending on the choice of the polynomial degree (e.g., 8000 samples). As such, to optimize the system throughput, it is highly required to queue data samples and process them as one data batch. Otherwise, the system performance degrades significantly if each individual data sample is processed separately. This constant overhead, in turn, bounds the applicability of CryptoNets in scenarios where data samples generated by distributed clients need to be processed with minimal latency.

This paper introduces DeepSecure, the first provably-secure framework for *scalable* DL-based analysis of data collected by distributed clients. DeepSecure enables applying the state-of-the-art DL models on sensitive data without sacrificing the accuracy to obtain security. Consistent with the literature, we assume an honest-but-curious adversary model for a generic case where both DL parameters and input data must be kept private. DeepSecure proposes the use of Yao's Garbled Circuit (GC) protocol to securely perform DL execution. In contrast with the prior work based on Homomorphic encryption [8], our methodology does not involve a trade-off between utility and privacy. We show that our framework is the best choice for scenarios in which the number of samples collected by each distributed client is less than 2600 samples; the clients send the data to the server for processing with the least possible delay.

Our approach is well-suited for streaming settings where clients need to dynamically analyze their data as it is collected over time without having to queue the samples to meet a certain batch size. In DeepSecure framework, we further provide mechanisms based on secret sharing to securely delegate GC computations to a third party for constrained embedded devices.

The function to be securely evaluated in GC should be represented as a list of Boolean logic gates (a.k.a., *netlist*). We generate the netlists required for deep learning using logic synthesis tools with GC-optimized custom libraries as suggested in [9]. DeepSecure leverages sequential circuit design to provide a set of scalable libraries and tools for deployment of different DL models with various sizes. Our custom synthesis library includes the first GC-optimized implementation of Tangent-Hyperbolic and Sigmoid functions used in DL. We also provide the enhanced implementation of matrix-vector multiplication to support signed input data as opposed to the realization reported in [9].

The computation and communication workload of GC protocol in DeepSecure framework is explicitly governed by the number of neurons in the target DL network and input data size. We further introduce a set of novel low-overhead pre-processing techniques to reduce data and DL network footprint without sacrificing neither the accuracy nor the data confidentiality. Our pre-processing approach is developed based on two sets of innovations: (i) transformation of input data to an ensemble of lower-dimensional subspaces, and (ii) avoiding the execution of neutral (inactive) neurons by leveraging the sparsity structure inherent in DL models. The explicit contributions of this paper are as follows:

- Proposing DeepSecure, the first provably-secure framework that simultaneously enables *accurate* and *scalable* privacy-preserving DL execution for distributed clients. Our approach is based on Yao's GC protocol and is amenable to embedded devices.
- Devising new custom libraries to generate *GC-optimized netlists* for required DL network computations using standard logic synthesis tools. The libraries include the first GC-optimized implementation of Tanh and Sigmoid functions.
- Incepting the idea of *data and DL network* pre-processing in the secure function evaluation settings. Our approach leverages the fine- and coarse-grained data and DL network parallelism to avoid unnecessary computation/communication in the execution of Yao's GC protocol.
- Introducing a *low-overhead secure outsourcing* protocol to provide support for constrained embedded platforms such as smartphones, medical implants, and wearable devices.
- Providing proof-of-concept evaluations of various visual, audio, and smart-sensing benchmarks. Our evaluations corroborate DeepSecure's scalability and practicability for distributed users compared with the HE-based solution.

## 2 PRELIMINARIES
### 2.1 Deep Learning Networks
DL refers to learning a hierarchical non-linear model that consists of several processing layers stacked on top of one the other. To perform data inference, the raw values of data features are fed into the first layer of the DL network known as the input layer. These raw features are gradually mapped to higher-level abstractions through the intermediate (hidden) layers of the DL model. The acquired data abstractions are then used to predict the label in the last layer of the DL network (a.k.a., the output layer).

Deep Neural Networks (DNNs) and Convolutional Neural Networks (CNNs) are the two main categories of neuron networks widely used in deep learning domain [1]. These two types of neural networks share many architectural similarities: CNNs are composed of additional convolution layers on top of fully-connected networks that form the foundation of DNNs. The use of convolutional layers in a CNN model makes them better-suited for interpreting data measurements with strong local connectivity (e.g., visual data), whereas DNNs pose a more generic architecture to model datasets which may not show a solid local dependency pattern (e.g., audio data). Table 1 outlines common hidden layers used in different DL networks.

The state of each neuron (unit) in a DL model is determined in response to the states of the units in the prior layer after applying a non-linear activation function. Commonly used activation functions for hidden layers include logistic sigmoid, Tangent-hyperbolic (Tanh), and Rectified Linear Unit (ReLu). The output layer is an exception for which a Softmax regression is typically used to determine the final inference. Softmax regression (or multinomial logistic regression) is a generalization of logistic regression that maps a $\mathcal{P}$-dimensional vector of arbitrary real values to a $\mathcal{P}$-dimensional vector of real values in the range of [0, 1). The final inference for each input sample can be determined by the output unit that has the largest conditional probability value [1].

Figure 1 demonstrates a schematic depiction of a CNN model consisting of conventional, pooling, fully-connected, and non-linearity layers. As we will explain in Section 4.2, each of these layers can be effectively represented as a Boolean circuit used in GC. An end-to-end DL model is formed by stacking different layers on top of each other. Note that many of the computations involved in DL inference, such as non-linearity layers, cannot be accurately represented by polynomials used in HE. For instance, approximating a Rectified Linear unit in HE requires using high-order polynomials, whereas a ReLu can be accurately represented by a Multiplexer in Boolean circuits.

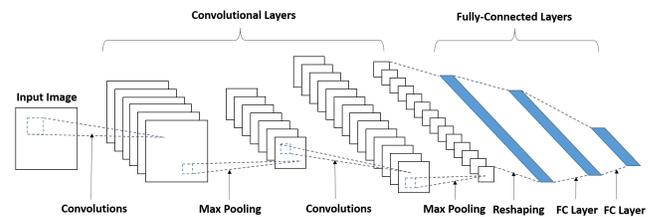

**Figure 1: Schematic depiction of a typical CNN model used for image classification [10].**

### 2.2 Cryptographic Protocols
In the following, we provide a brief description of the cryptographic protocols used in this paper.

Table 1: Commonly used layers in DNN and CNN models.

| DL Layer | Description | Computation | |
|---|---|---|---|
| **2D Convolution** (C) | Multiplying the filter parameters ($\theta_{ij}$) with the post-nonlinearity values in the preceding layer ($z_{ij}^{(l-1)}$) and summing up the results | $x_{ij}^{(l)} = \sum_{s_1=1}^{k} \sum_{s_2=1}^{k} \theta_{s_1 s_2}^{(l-1)} \times z_{(i+s_1)(j+s_2)}^{(l-1)}$ | Weighted Sum |
| **Max Pooling** ($M_1P$) | Computing the maximum value of $k \times k$ overlapping regions of the preceding layer | $x_{ij}^{(l)} = Max(z_{(i+s_1)(j+s_2)}^{(l-1)})\ s_1 \in \{1, 2, ..., k\}$ $s_2 \in \{1, 2, ..., k\}$ | Maximum |
| **Mean Pooling** ($M_2P$) | Computing the mean value of $k \times k$ non-overlapping regions of the preceding layer | $x_{ij}^{(l)} = Mean(z_{(i+s_1)(j+s_2)}^{(l-1)})\ s_1 \in \{1, 2, ..., k\}$ $s_2 \in \{1, 2, ..., k\}$ | Mean |
| **Fully-Connected** (FC) | Multiplying the parameters of the $l^{th}$ layer ($\theta_{ij}^{l}$) with the post-nonlinearity values in the preceding layer ($z_i^{l-1}$) | $x_i^{(l)} = \sum_{j=1}^{n_{l-1}} \theta_{ij}^{(l-1)} \times z_j^{(l-1)}$ | Matrix-Vector Multiplication |
| **Non-Linearity** (NL) | Softmax | $z_i^{(l)} = \frac{e^{x_i^{(l)}}}{\sum_{j=0}^{n_l} e^{x_j^{(l)}}}$ | CORDIC |
| | Sigmoid | $z_i^{(l)} = \frac{1}{1+e^{-x_i^{(l)}}}$ | CORDIC |
| | Tangent-Hyperbolic (Tanh) | $z_i^{(l)} \frac{Sinh(x_i^{(l)})}{Cosh(x_i^{(l)})}$ | CORDIC |
| | Rectified Linear unit (ReLu) | $z_i^{(l)} = Max(0, x_i^{(l)})$ | Maximum |

### 2.2.1 Oblivious Transfer

Oblivious Transfer (OT) is a cryptographic protocol that runs between a Sender (S) and a Receiver (R). The receiver R obliviously selects one of potentially many pieces of information provided by S. Particularly, in a 1-out-of-n OT protocol, the sender S has $n$ messages ($x_1, ..., x_n$) and the receiver R has an index $r$ where $1 \leq r \leq n$. In this setting, R wants to receive $x_r$ among the sender's messages without the sender learning the index $r$, while the receiver only obtains one of the $n$ possible messages [11].

### 2.2.2 Garbled Circuit

Yao's garbled circuit protocol [12] is a cryptographic protocol in which two parties, Alice and Bob, jointly compute a function $f(x, y)$ on their inputs while keeping the inputs fully private. In GC, the function $f$ should be represented as a Boolean circuit with 2-input gates (e.g., XOR, AND, etc.). The input from each party is represented as input wires to the circuit. All gates in the circuit have to be topologically sorted which creates a list of gates called *netlist*. GC has four different stages: (i) garbling which is only performed by Alice (a.k.a., Garbler). (ii) Data transfer and OT which involves both parties, Alice and Bob. (iii) Evaluating, only performed by Bob (a.k.a., Evaluator), and finally (iv) merging the results of the first two steps by either of the parties.

**(i) Garbling.** Alice garbles the circuit by assigning two random $k$-bit labels to each wire in the circuit corresponding to semantic values one and zero ($k$ is the security parameter usually set to 128). For instance, for input wire number 5, Alice creates 128-bit random string $l_5^0$ as a label for semantic value zero and $l_5^1$ for semantic value one. For each gate, a garbled table is computed. The very first realization of the garbled table required four different rows, each corresponding to one of the four possible combinations of inputs labels. Each row is the encryption of the correct output key using two input labels as the encryption key [13]. As an example, assume wire 5 ($w_5$) and 6 ($w_6$) are input to an XOR gate and the output is wire 7 ($w_7$). Then, the second row of the garbled table which corresponds to ($w_5 = 0$) and ($w_6 = 1$) is equivalent to $Enc_{(l_5^0, l_6^1)}(l_7^1)$. To decrypt any garbled table, one needs to possess the associated two input labels. Once Garbler creates all garbled tables, the protocol is ready for the second step.

**(ii) Transferring Data and OT.** In this step, Alice sends all the garbled tables along with the correct labels corresponding to her actual input to Bob. For instance, if the input wire 8 belongs to her and her actual input for that wire is zero, she sends $l_8^0$ to Bob. In order for Bob to be able to decrypt and evaluate the garbled tables (step 3), he needs the correct labels for his input wires as well. This task is not trivial nor easy. On the one hand, Bob cannot send his actual input to Alice (because it undermines his input privacy). On the other hand, Alice cannot simply send both input labels to Bob (because Bob can then infer more information in step 3). To effectively perform this task, OT protocol is utilized. For each input wire that belongs to Bob, both parties engage in a 1-out-of-2 OT protocol where the selection bit is Bob's input and two messages are two labels from Alice. After all required information is received by Bob, he can start evaluating the garbled circuit.

**(iii) Evaluating.** To evaluate the garbled circuit, Bob starts from the first garbled table and uses two input labels to decrypt the correct output key. All gates and their associated dependencies are topologically sorted in the netlist. As such, Bob can perform the evaluation one gate at a time until reaching the output wires without any halts in the process. In order to create the actual plaintext output, both the output mapping (owned by Alice) and final output labels (owned by Bob) are required; thereby, one of the parties, say Bob, needs to send his share to the other party (Alice).

**(iv) Merging Results.** At this point, Alice can easily compute the final results. To do so, she uses the mapping from output labels to the semantic value for each output wire. The protocol can be considered finished after merging the results (as in DeepSecure) or Alice can also share the final results with Bob.

## 2.3 Garbled Circuit Optimizations

In Section 2.2.2, we have described the GC protocol in its simplest form. During the past decade, several optimization methodologies have been suggested in the literature to minimize the overhead of executing GC protocol. In the following, we summarize the most important optimizations techniques that we also leverage to achieve an efficient deployment of DeepSecure framework.

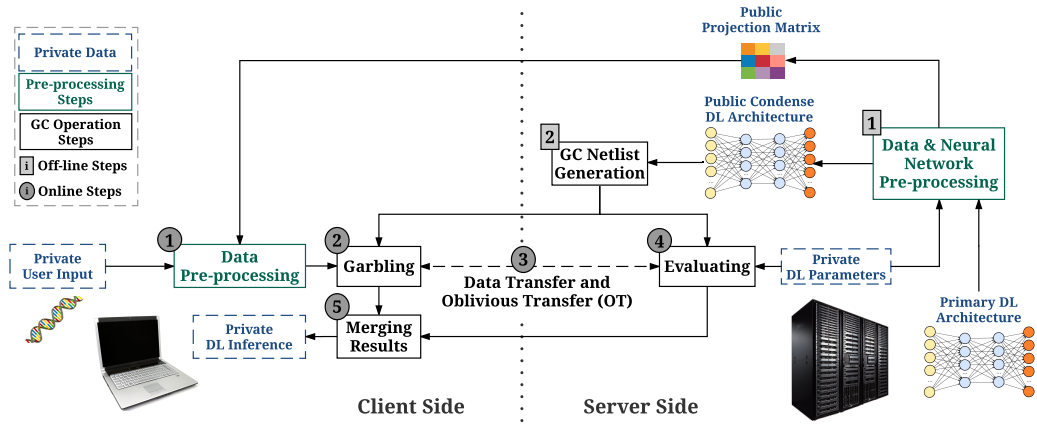

**Figure 2: Global flow of DeepSecure framework including both off-line (indexed by rectangular icons) and online (indexed by oval icons) steps. The operations shown in the left hand side of the figure are executed by the client (Alice) while the operations on the right hand side are performed by the server (Bob).**

**Row-Reduction:** As we discussed earlier, the initial garbled table consists of four rows. Authors in [14] proposed a technique to reduce the number of rows in the garbled table to three. Since the main portion of communication is to transfer the garbled tables, this technique results in almost 25% reduction in communication.

**Free-XOR:** Perhaps one of the most important optimizations of GC is Free-XOR [15]. The Free-XOR methodology enables the evaluation of XOR, XNOR, and NOT gates without costly cryptographic encryption. Therefore, it is highly desirable to minimize the number of non-XOR gates in the deployment of the underlying circuit.

**Half-Gates:** This technique that is proposed in [16], further reduces the number of rows for AND gates from three to two, resulting in 33% less communication on top of the Row-reduction optimization.

**Garbling with Fixed-Key Block Cipher:** This methodology [13] introduces an encryption mechanism for garbled tables based on fixed-key block ciphers (e.g., AES). Many of the modern processors have AES-specific instructions in their Instruction Set Architecture (ISA) which, in turn, makes the garbling and evaluating process significantly faster using the fixed-key cipher.

**Sequential Garbled Circuit:** For years the GC protocol could have only been used for Combinational circuits (a.k.a., acyclic graphs of gates). Authors in [9] suggested a new approach that enables garbling/evaluating sequential circuits (cyclic graphs). In their framework, one can garble/evaluate a sequential circuit iteratively for multiple clock cycles. Following the approach presented in [9], the function $f$ can be described with a Hardware Description Language (HDL) and compiled with a logic synthesis tool.

## 2.4 Security Model

We assume an *Honest-but-Curious* (HbC) security model in which the participating parties follow the protocol they agreed on, but they may want to deduce more information from the data at hand. We focus on this security model because of the following reasons:

- HbC is a standard security model in the literature [13, 17] and is the first step towards stronger security models (e.g., security against malicious adversaries). Our solution can be readily modified to support *malicious* models by following the methodologies presented in [18–21]. Note that stronger security models rely on multiple rounds of HbC with varying parameters. As such, the efficiency of DeepSecure is carried out to those models as well.
- Many privacy-preserving DL execution settings naturally fit well in HbC security. For instance, when all parties have the incentive to produce the correct result (perhaps when the DL inference task is a paid service). In these settings, both data provider and the server that holds the DL model will follow the protocol in order to produce the correct outcome.

## 3 DeepSecure FRAMEWORK

Figure 2 demonstrates the overall flow of DeepSecure framework. DeepSecure consists of two main components to securely perform data inference in the context of deep learning: (i) GC-optimized execution of the target DL model (Section 3.1), and (ii) data and DL network transformation (Section 3.2).

### 3.1 DeepSecure GC Core Structure

Figure 3 illustrates the core structure of DeepSecure framework. In our target setting, a cloud server (Bob) holds the DL model parameters trained for a particular application, and a delegated client (Alice) owns a data sample for which she wants to securely find the corresponding classified label (a.k.a., inference label).

DL models have become a well-established machine learning technique. Commonly employed DL topologies and cost functions are well-known to the public. Indeed, what needs to be kept private from the cloud server's perspective is the DL model parameters that have been tuned for a particular task using massive statistical databases and devoting large amounts of computing resources for several weeks/months. Data owners, on the other hand, tend to leverage the models already tuned by big technology companies to figure out the inference label for their private data while keeping their data fully private.

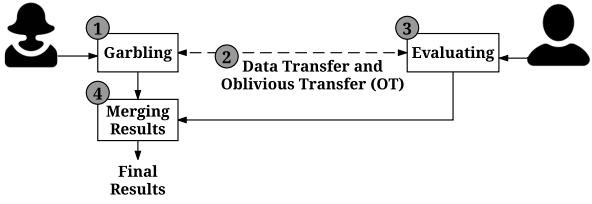

Figure 3: DeepSecure GC core structure.

Table 2: GC Computation and Communication Costs for realization of a DNN model.

| Computation and Communication Costs |
|---|
| $T_{comp} = \beta_{mult} \sum_{l=1}^{n_l-1} n^{(l)} n^{(l+1)} + \beta_{add} \sum_{l=2}^{n_l} n^{(l)} + \beta_{act} \sum_{l=2}^{n_l} n^{(l)}$ |
| $\beta_{opr} = \frac{N_{opr}^{XOR} \times C_{opr}^{XOR} + N_{opr}^{Non\_XOR} \times C_{opr}^{Non\_XOR}}{f_{CPU}}$ |
| $n_l$: total number of DL layers |
| $\beta_{mult}$: computational cost of a multiply operation in GC protocol |
| $\beta_{add}$: computational cost of an add operation in GC protocol |
| $\beta_{act}$: computational cost of a non-linearity operation in GC protocol |
| $N^{XOR}$: number of XOR gates |
| $N^{Non\_XOR}$: number of non_XOR gates |
| $C^{XOR}$: garbling/evaluating cost of a XOR gate |
| $C^{Non\_XOR}$: garbling/evaluating cost of a Non_XOR gate |
| $f_{CPU}$: CPU clock frequency |
| $T_{comm} = \frac{\alpha_{mult} \sum_{l=1}^{n_l-1} n^{(l)} n^{(l+1)} + \alpha_{add} \sum_{l=2}^{n_l} n^{(l)} + \alpha_{act} \sum_{l=2}^{n_l} n^{(l)}}{BW_{net}}$ |
| $\alpha_{opr} = N_{opr}^{Non\_XOR} \times 2 \times N_{bits}$ |
| $BW_{net}$: operational communication bandwidth |
| $\alpha_{mult}$: communication cost of a multiply operation in GC protocol |
| $\alpha_{add}$: communication cost of an add operation in GC protocol |
| $\alpha_{act}$: communication cost of a non-linearity operation in GC protocol |
| $N_{bits}$: GC security parameter |

DeepSecure enables computing the pertinent data inference label in a provably-secure setting while keeping both the DL model's parameters and data sample private. To perform a particular data inference, the netlist of the publicly known DL architecture[1] should be generated prior to the execution of the GC protocol. The execution of the GC protocol involves four main steps per Section 2.2.2: (i) the client (data owner) garbles the Boolean circuit of the DL architecture. (ii) The client sends the computed garbled tables from the first step to the cloud server along with her input wire labels. Both client and the cloud server then engage in a 1-out-of-2 Oblivious Transfer (OT) [11] protocol to obliviously transfer the wire labels associated with cloud server's inputs. (iii) The cloud server evaluates (executes) the garbled circuit and computes the corresponding encrypted data inference. (iv) The encrypted result is sent back to the client to be decrypted using the garbled keys so that the true inference label is revealed.

#### 3.1.1 GC Communication and Computation Overhead
Table 2 details the computation and communication cost for execution of a fully-connected DNN. A similar setup applies to the CNN models in which a set of convolutions are performed per layer. DeepSecure finds an estimation of the physical coefficients ($\beta$ and $\alpha$) listed in Table 2 by running a set of subroutines as we will discuss in Section 4.3. The total communication needed between client and the server is proportional to the number of non-XOR gates since only the garbled tables for non-XOR gates need to be transferred.

As shown in Table 2, the computation and communication overhead of DL execution using GC protocol is explicitly governed by the number of neurons (units) per DL layer. As such, we suggest a set of data and DL network transformation as an arbitrarily pre-processing step to reduce the computation and communication overhead of GC protocol for DL inference.

### 3.2 Data and DL Network Pre-processing
DeepSecure pre-processing consists of two main steps: (i) data projection (Section 3.2.1), and (ii) DL network distillation (Section 3.2.2).

#### 3.2.1 Data Projection
The input layer size of a neural network is conventionally dictated by the feature space size of the input data samples. Many complex modern data matrices that are not inherently low-rank can be modeled by a composition of multiple lower-rank subspaces. This type of composition of high-dimensional dense (non-sparse but structured) data as an ensemble of lower dimensional subspaces has been used earlier by data scientists and engineers to facilitate knowledge extraction or achieve resource efficiency [22–25]. DeepSecure, for the first time, introduces, implements, and automates the idea of pre-processing and systematic alignment of these data subspaces as a way to achieve performance optimization for GC execution of a DL model.

---
[1] DL architecture refers to the number and type of layers and not the values of the pertinent private DL parameters.

As we experimentally demonstrate in Section 4, the main bottleneck in the execution of GC protocol is the communication overhead between the server (Bob) and client (Alice). Therefore, we focus our pre-processing optimization to minimize the GC communication workload ($T_{comm}$) customized to the application data and DL model (see Table 2 for the characterization of communication overhead in accordance with the data and DL network size). In order to perform data pre-processing in DeepSecure framework, the server (Bob) needs to re-train its private DL parameters according to the following objective function:

$$\begin{aligned} \underset{D,\,\tilde{DL}_{param}}{Minimize} \ (T_{comm}) \ s.t.,\ ||A - DC||_F &\leq \epsilon ||A||_F \\ \delta(\tilde{DL}_{param}) &\leq \delta(DL_{param}) \\ l &\leq m, \end{aligned} \quad (1)$$

where $A_{m \times n}$ is the raw input training data (owned by the server) that we want to factorize into a *dictionary matrix* $D_{m \times l}$ and a low-dimensional *data embedding* $C_{l \times n}$ that is used to re-train the DL model. Here, $\delta(.)$ is the partial validation error corresponding to the pertinent DL parameters. We use $DL_{param}$ to denote the initial DL parameters acquired by training the target DL model using raw data features $A$. Whereas, $\tilde{DL}_{param}$ indicates the updated parameters after re-training the underlying DL model using the projected data embedding $C$. $||\cdot||_F$ denotes the Frobenius norm and $\epsilon$ is an intermediate approximation error that casts the rank of the input data.

**Algorithm 1** Off-line re-training process locally performed by server (Bob)

**Input:** Training data ($A$), batch size ($n_{batch}$), training labels ($L_{Tr}$), projection threshold ($\gamma$), number of samples to be evaluated before early stopping (*patience*), and pre-trained DL parameters ($DL_{param}$).

**Output:** Updated DL parameters $\tilde{DL}_{param}$, and projection matrix $W$.

1: $D \leftarrow$ empty
2: $C \leftarrow$ empty
3: $\tilde{DL}_{param} \leftarrow DL_{param}$
4: $N_{Tr} \leftarrow |A|$ //Number of columns in A
5: $\delta_{best} \leftarrow 1$
6: $\delta \leftarrow 1$
7: $itr \leftarrow 0$
8: $j \leftarrow 0$
9: $i \leftarrow 0$
10: **while** $i \leq N_{Tr}$ **do**
11: $\quad C_i \leftarrow 0$ //Column i in matrix C
12: $\quad$ **if** $D$ is empty **then**
13: $\quad\quad V_p(a_i) = 1$
14: $\quad$ **else**
15: $\quad\quad V_p(a_i) = \frac{||D(D^TD)^{-1}D^Ta_i - a_i||_2}{||a_i||_2}$
16: $\quad$ **end if**
17: $\quad$ **if** $\delta \leq \delta_{best}$ **then**
18: $\quad\quad \delta_{best} \leftarrow \delta$
19: $\quad\quad itr \leftarrow 0$
20: $\quad$ **else**
21: $\quad\quad itr \leftarrow itr + 1$
22: $\quad$ **end if**
23: $\quad$ **if** $V_p(a_i) > \gamma$ **&** $itr <$ *patience* **then**
24: $\quad\quad D \leftarrow [D, a_i/\sqrt{||a_i||_2}]$
25: $\quad\quad C_{ij} = \sqrt{||a_i||_2}$ //Element j in column i
26: $\quad\quad j \leftarrow j + 1$
27: $\quad$ **else**
28: $\quad\quad C_i \leftarrow D(D^TD)^{-1}D^Ta_i$
29: $\quad$ **end if**
30: $\quad C \leftarrow [C, C_i]$
31: $\quad i \leftarrow i + 1$
32: $\quad$ **if** $i \mod n_{batch} == 0$ **then**
33: $\quad\quad \tilde{DL}_{param} \leftarrow UpdateDL(\tilde{DL}_{param}, C, L_{Tr})$
34: $\quad\quad \delta \leftarrow UpdateValidationError(\tilde{DL}_{param})$
35: $\quad$ **end if**
36: **end while**
37: $W = D(D^TD)^{-1}D^T$

Algorithm 1 outlines the pseudocode of our data pre-processing step performed by the server (Bob). To solve Eq. 1, we first initiate the matrices $D$ and $C$ as empty sets. DeepSecure gradually updates the corresponding data embeddings by streaming the input training data. In particular, for a batch of training data samples ($A_i$), we first calculate a projection error, $V(A_i)$, based on the current values of the dictionary matrix $D$. This error shows how well the newly added samples can be represented in the space spanned by $D$. If the projection error is less than a user-defined threshold ($\gamma$), it means the current dictionary $D$ is good enough to represent those new samples ($A_i$). Otherwise, the corresponding data embeddings are modified to include the new data structure imposed by the new batch of data. The data embedding $C$ is then used to update previous DL parameters using multiple rounds of forward and backward propagation (Line 33 Algorithm 1).

Once the DL model is re-trained using the projected data embedding $C$, the underlying projection matrix ($W = DD^+$)[2] is publicly released to be used by the clients during DL execution phase. As we will discuss in Section 3.7, $W$ does not reveal any information regarding the training data nor the DL parameters. We emphasize that re-training of conventional DL model is a *one-time off-line* process performed by the server. As such, the data pre-processing overhead during GC execution only involves a matrix-vector multiplication, $Y_i = WX_i$, that is performed prior to garbling on the client side (Algorithm 2). Here, $X_i$ indicates the raw data owned by the client (Alice), $W$ denotes the projection matrix, and $Y_i$ is the projected data in the space spanned by columns of matrix $W$.

**Algorithm 2** Online data pre-processing performed by the client (Alice)

**Input:** Raw data measurement ($X$), projection matrix $W$, number of client's samples $N_{client}$.

**Output:** Projected data samples $Y$.

1: $i \leftarrow 0$
2: **while** ($i < N_{client}$) **do**
3: $\quad Y_i \leftarrow WX_i$
4: $\quad i \leftarrow i + 1$
5: **end while**

### 3.2.2 DL Network Pre-processing

Recent theoretical and empirical advances in DL has demonstrated the importance of sparsity in training DL models [26–28]. Sparsity inducing techniques such as rectifying non-linearities and $\mathcal{L}_1$ penalty are key techniques used to boost the accuracy in training neural networks with millions of parameters.

To eliminate the unnecessary garbling/evaluation of non-contributing neurons in a DL model, we suggest pruning the underlying DL network prior to netlist generation for the GC protocol. In our DL network pre-processing, the connections with a weight below a certain threshold are removed from the network. The condensed network is re-trained as suggested in [28] to retrieve the accuracy of the initial DL model. DeepSecure network pre-processing step is a one-time off-line process performed by the server (Bob). Our approach is built upon the fact that DL models are usually over-parameterized and can be effectively represented with a sparse structure without a noticeable drop in the pertinent accuracy.

---
[2] $D^+$ indicates the pseudo-inverse of the matrix $D$.

Note that using conventional GC protocol, it is not feasible to skip the multiplication/addition in evaluating a particular neural (unit) in a DL model. Our network pre-processing cuts out the non-contributing connections/neurons per layer of a DL network. It, in turn, enables using the sparse nature of DL models to significantly reduce the computation and communication workload of executing the GC protocol. The off-line step 1 indicated in Figure 2 corresponds to both data pre-processing outlined in Algorithm 1 and neural network pruning.

### 3.3 Secure Outsourcing

DeepSecure provides support for secure outsourcing of the GC protocol to a proxy cloud server. This setting is well-suited for users with severe resource constraints who may not be able to perform the garbling task in a reasonable amount of time. We assume the same level of security for the proxy server, being honest-but-curious. This means that similar to the main server, we do not trust the proxy server but we expect it to follow the protocol and do not collude with the main server. Figure 4 illustrates the overall flow of DeepSecure framework in the secure outsourcing scenario. The proxy server can be a simple personal computer that is connected to the Internet.

Our secure outsourcing scheme is based on XOR-sharing technique and works as follows: Client needs to generate $l$-bit random string $s$, where $l$ is the bit-length of her input ($x$). Then, she XORs her input with $s$, resulting in $x \oplus s$. She sends $s$ to one of the servers and $x \oplus s$ to the other one. The Boolean circuit that is garbled in the GC protocol remains the same except that it has one layer of XOR gates at the initial stage of the circuit to generate the true input $x$ securely inside the circuit (($x \oplus s) \oplus s = x$).

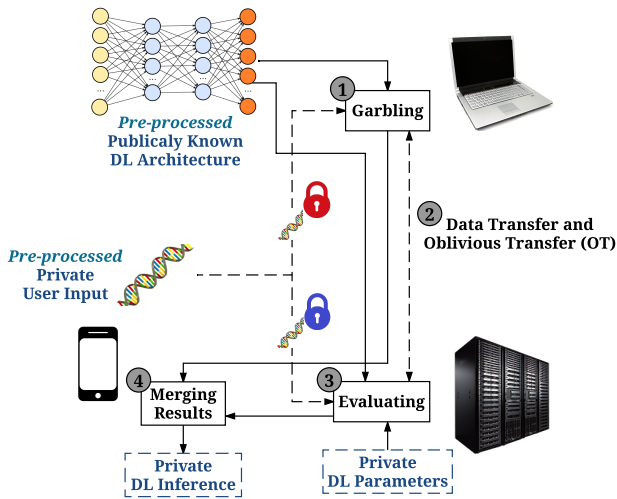

Figure 4: The overall flow of DeepSecure framework in the resource-constrained settings.

**Communication and Computation Overhead.** The secure outsourcing model does not introduce significant extra overhead. In particular, users only need to generate a random string and do a simple XOR operation. The GC protocol that runs between two servers is the same as the original protocol except that we need to put one layer of XOR gates in front of the circuit to produce the true input ($x$). Thanks to the Free-XOR technique [15], this additional layer is almost free of charge resulting in almost the same computation and communication overhead as the original scheme.

### 3.4 GC-Optimized Circuit Components Library

As we explained earlier, the GC protocol requires the function of interest being represented as a Boolean circuit. Following the Free-XOR optimization [15], the XOR gates are almost free of cost and the garbled table needs to be generated and transferred only for the non-XOR gates. Therefore, to optimize the computation and communication costs, one needs to minimize the number of non-XOR gates. We leverage the industrial synthesis tool to optimize the resulting netlist by setting the area overhead for XOR gates to zero and for all the other non-XOR gates to one. As such, forcing the synthesis tools to create a netlist with minimum area, outputs a netlist with least number of non-XOR gates.

Our custom synthesis library includes the GC-optimized realization of all the necessary computation modules in a neural network (Table 1). We design optimized fundamental blocks e.g., multiplexer (MUX), comparator (CMP), adder (ADD), and multiplier (MULT) such that they incur the least possible non-XOR gates. We add our optimized blocks to the library of the synthesis tool. The detail results of our GC-optimized realization of DL layers are presented in Section 4.2.

### 3.5 GC Memory Footprint

We use sequential circuits as opposed to combinational circuits traditionally used in GC for generating the netlists. Sequential circuits are compact and can be scaled up easily due to their low memory footprint. Therefore, the underlying circuit can be folded into a more compact representation and be run for multiple clock cycles. For instance, instead of instantiating all the MULT and ADD modules used in the matrix multiplication individually, we can use one MULT, one ADD, and multiple registers to accumulate the result. A single multiplication is performed at a time and the result is added to the previous steps, continuing until all operations are done. As such, the circuit generation is no longer the bottleneck, as the memory footprint of the circuit will be constant while the circuit needs to be run for multiple iterations.

### 3.6 DeepSecure Modular Structure

Each DL layer outlined in Table 1 is implemented as a single module in DeepSecure framework. These layers can be easily connected to one another in order to form an end-to-end DL network. The *modular nature* of DeepSecure enables users to create different DL models with arbitrary architectures. For instance, a user can instantiate one layer of convolutional layer followed by a non-linear activation function and max pooling by easily stacking one layer on the top of the previous layer. We have benchmarked multiple well-known DL models including Convolutional Neural Networks and Deep Neural Networks, each of which is explained in detail in Section 4.5. We leverage Fixed-point number format for our evaluations presented in Section 4.2. However, we emphasize that our GC-optimized library also provides support for Floating-point accuracy as well.

## 3.7 Security Proof

In this section, we provide a comprehensive security proof of DeepSecure in the Honest-but-Curious (HbC) adversary model. Our core secure function evaluation engine is the GC protocol. GC is proven to be secure in HbC adversary model [29]; thereby, any input from either client or server(s) to GC will be kept private during the protocol execution. The Garbled circuit optimization techniques (Section 3.4) that we leverage in DeepSecure to reduce the number of non-XOR gates of circuit components do not affect the security of our framework. This is because the security of the GC protocol is independent of the topology of the underlying Boolean circuit. As such, we only need to provide the security proof of the three modifications that are performed outside of the GC protocol: (i) data pre-processing, (ii) DL network pre-processing, and (iii) the two secret shares generated in secure outsourcing mode that are released to the two servers (Section 3.3). In the following, we provide detailed proofs for each of aforementioned modifications.

**(i) Data Pre-processing.** In step one of off-line processing depicted in Figure 2, the projection matrix ($W$) is computed and released publicly. In Proposition 3.1, we provide the security guarantees for this step.

**Proposition 3.1.** *Projection Matrix ($W$) reveals nothing but the subspace of dictionary matrix ($D$) from which the matrix $D$ cannot be reconstructed.*

**Proof:** Let $D_{m\times l} = U_{m\times r} \times \Sigma_{r\times r} \times V^T_{r\times l}$ denote the Singular Value Decomposition (SVD) of the dictionary matrix $D$, where $r$ is the rank of $D$ ($r = min(m,l)$). As such,

$$\begin{aligned}
W &= DD^+ \\
&= D(D^T D)^{-1} D^T \\
&= U\Sigma V^T (V\Sigma^T U^T U\Sigma V^T)^{-1} V\Sigma^T U^T \\
&= U\Sigma (V^T V^{T-1})(\Sigma^T I_r \Sigma)^{-1} (V^{-1} V)\Sigma^T U^T \\
&= U\Sigma I_r (\Sigma^T \Sigma)^{-1} I_r \Sigma^T U^T \\
&= UU^T,
\end{aligned} \quad (2)$$

where $I_r$ indicates the identity matrix of size $r \times r$. As such, the projection matrix $W$ does not reveal any information regarding the actual values of the dictionary matrix $D$ but the subspace spanned by its column space $U$ (a.k.a., *left-singular vectors*). Note that for a given set of left-singular vectors $U$, there exist infinite possible data matrices that reside in the same column space. As such, the dictionary matrix $D$ cannot be reconstructed without having access to corresponding right-singular vectors $V$ and the singular value set $\Sigma$. If revealing the subspace spanned by $U$ is not tolerable by the server, this pre-processing step can be skipped.

**(ii) DL Network Pre-processing.** In this pre-processing stage, the inherent sparsity of the neural network is utilized to produce a new model which requires less computation. In order to avoid garbling/evaluating unnecessary components in the Boolean circuit, the server needs to modify the netlist used in the GC protocol. Therefore, the sparsity map of the network is considered as a public knowledge and will be revealed to the Garbler as well (Garbler needs to garble the circuit based on the netlist). However, the sparsity map only contains information regarding which part of the network does not contribute to the output and reveals nothing about the private network parameters. Just like the data pre-processing step, if revealing the sparsity map is not acceptable by the server (DL model owner), this step can be skipped.

**(iii) Outsourcing.** As discussed in Section 3.3, to securely outsource the computation of the GC protocol to an *untrusted* server Alice needs to generate a random number, send it to server one, XOR her input with the random string, and send the XOR results to the second server. This approach is secure based on the following proposition.

**Proposition 3.2.** *The XOR-sharing technique is secure with untrusted servers as long as two servers do not collude.*

**Proof:** The first server only receives a purely random string which is totally independent of the Client's input data. Server one should input this data to the GC protocol (considering that each party must follow the protocol in HbC adversary model). What the second server receives, is equivalent to One Time Pad (OTP) encryption of Client's input (using the random pad that is sent to the first server). Server two also should input this data to the GC protocol. Since OTP is proven to be secure [30], the only possible way to decrypt the Client's input is to have both the random pad and the encrypted data which is not possible due to the non-colluding assumption.

# 4 EVALUATIONS

## 4.1 Setup Experiment

We use Synopsys Design Compiler 2010.03-SP4 to generate the Boolean circuits. The timing analysis is performed by two threads on an Intel Core i7-2600 CPU working at 3.4$GHz$ with 12$GB$ RAM and Ubuntu 14.04 operating system. We quantify the operational throughput in our setting in Section 4.4. In all of the experiments, the GC security parameter is set to 128-bit.

## 4.2 Circuit Synthesis

We implement each of the functions required for DL evaluation (Table 1) using Verilog HDL and compile them with Synopsys Design Compiler [31]. The results of our synthesized circuits in terms of the number of XOR and non-XOR are summarized in Table 3. Table 3 includes three different approximation methods for Tanh and Sigmoid. The error shown in this table reflects the level of approximation in the realization of different variants of a particular function and not the representational error. In any digital representation with $b$ fractional bits, there is an additional error of less than or equal to $2^{-(b+1)}$. This error is introduced as a result of truncating the precise representation which is inescapable in any digital system. In our evaluations, we use 1 *bit* to hold the sign, 3 *bits* for integer part, and 12 *bits* for the fractional part (denoted as $b$). Therefore, a representational error of $2^{-13}$ is present throughout the entire network. An error rate of zero in Table 3 means the result of the computation is precise up to $b$ fractional bits. Any non-zero error in the Table 3 implies that the computation will add the specified amount of error to the final result.

**Table 3: Number of XOR and non-XOR gates for each element of DL networks.**

| Name | #XOR | #Non-XOR | Error |
|---|---|---|---|
| $Tanh_{LUT}$ | 692 | 149745 | 0 |
| $Tanh_{2.10.12}$ | 3040 | 1746 | 0.01% |
| $Tanh_{PL}$ | 5 | 206 | 0.22% |
| $Tanh_{CORDIC}$ | 8415 | 3900 | 0 † |
| $Sigmoid_{LUT}$ | 553 | 142523 | 0 |
| $Sigmoid_{3.10.12}$ | 3629 | 2107 | 0.04% |
| $Sigmoid_{PLAN}$ | 1 | 73 | 0.59% |
| $Sigmoid_{CORDIC}$ | 8447 | 3932 | 0 |
| ADD | 16 | 16 | 0 |
| MULT | 381 | 212 | 0 |
| DIV | 545 | 361 | 0 |
| ReLu | 30 | 15 | 0 |
| $Softmax_n$ | $(n-1) \cdot 48$ | $(n-1) \cdot 32$ | 0 |
| $A_{1 \times m} \cdot B_{m \times n}$ | $397 \cdot m \cdot n - 16 \cdot n$ | $228 \cdot m \cdot n - 16 \cdot n$ | 0 |

† To achieve k bit precision, CORDIC has to be evaluated for k iterations.

The Sigmoid function has a symmetry point at (0, 0.5). Therefore, one only needs to compute the function for one-half of the x-y pairs ($y_{x<0} = 1 - y_{x>0}$). Similarly, Tanh is an odd function, thus, we have $y_{x<0} = -y_{x>0}$. As shown in the Table 3, each approximation method results in a different error rate and circuit size. In the first Tanh realization, this function is computed using Look-Up-Table (LUT) which incurs a zero computational error. In the second realization, $Tanh_{2.10.12}$, we eliminate the effect of the two least significant fractional bits and the most significant integer bit of the input value $x$. In particular, we set Tanh(x) for any $x$ value greater than four, equal to one. This approximation has an error rate of 0.01%. Another less accurate but less costly approach is to approximate Tanh with a piece-wise linear function (denoted as $Tanh_{PL}$ in Table 3). In this realization, we estimate Tanh with seven different lines for $x >= 0$. $Tanh_{PL}$ is almost 700 times less costly than $Tanh$ with an error rate of 0.22%. Equivalently, we present three variants of Sigmoid function where $Sigmoid_{PLAN}$ is a piece-wise linear implementation for approximate computation of Sigmoid [32].

To compute DL non-linearity functions, we also evaluate COordinate Rotation DIgital Computer (CORDIC) circuit. Each iteration of computing CORDIC improves the final accuracy by one bit. As such, in order to achieve 12 bit accuracy, we need to iteratively evaluate the circuit 12 times. To operate CORDIC in hyperbolic mode, one needs to evaluate iterations $(3 \times i + 1)$ twice, which in turn, results in an overall 14 iterations per instance computation. CORDIC outputs Cosine-Hyperbolic (Cosh) and Sine-Hyperbolic (Sinh). We use these outputs to compute the corresponding values of Tanh ($\frac{Sinh(x)}{Cosh(x)}$) and Sigmoid function ($\frac{1}{1+Cosh(x)-Sinh(x)}$). The synthesized result provided in Table 3 shows the total number of gates for 14 iterations of evaluation plus one DIV operation for $Tanh_{CORDIC}$ with an additional two ADD operations for Sigmoid computation.

Softmax is a monotonically increasing function. Therefore, applying this function to a given input vector does not change the index of the maximum value (inference label index). As such, we use optimized CMP and MUX blocks to implement Softmax in DeepSecure framework. We provide different circuits for computing DL non-linear activation functions to offer speed/accuracy trade-off. One can choose each circuit according to her application criteria. We utilize the CORDIC-based implementation of Tanh and Sigmoid in our experiments presented in Section 4.5.

### 4.3 Performance Cost Characterization

Here, we quantify the GC performance parameters outlined in Section 3.1.1. In DeepSecure framework, garbling/evaluating each non-XOR and XOR gate requires 164 and 62 CPU clock cycles (*clks*) on average, respectively. As a result, the computational time for each operation (e.g., multiplication) can be defined as:

$$\beta_{opr} = \frac{N_{opr}^{XOR} \times 62 + N_{opr}^{non-XOR} \times 164}{f_{CPU}} \ sec, \quad (3)$$

where $f_{CPU}$ is the CPU clock frequency.

In GC, the garbled tables for each non-XOR gate need to be communicated between the two parties. Each table has two rows and each row is 128 bits. Therefore, the total amount of data that client has to send to the server (or the server-to-server communication overhead in secure outsourcing setting) can be defined as the following:

$$\alpha_{opr} = N_{opr}^{non-XOR} \times 2 \times 128 \ bit. \quad (4)$$

Total execution time is dominated by the time required to transfer the garbled tables as shown in Table 4.

### 4.4 GC Execution in Practice

Figure 5 illustrates the timing diagram of different GC steps for a sequential circuit. In the case of a combinational circuit, only one clock cycle of the operations will be executed. As we explained in Section 2.2.2, the process starts by Alice who garbles the circuit. After garbling, both parties engage in the OT protocol to obliviously transfer the labels. Garbled tables are then sent to Bob. At this point, while Bob starts evaluating the circuit, Alice starts garbling for the second clock cycle. Therefore, the total execution time of the protocol is not the summation of the execution time of both parties. This process continues until the circuit is garbled for specified number of clock cycles after which the result is sent back to Alice to decode the final data inference result. The garbling/evaluating time in GC is proportional to the total number of gates in the circuit, whereas the data transfer time mainly depends on the total number of non-XOR gates in the circuit. The OT execution overhead particularly depends on the input data size (number of input bits). In our target DL setting, DeepSecure achieves an effective throughput of $2.56M$ and $5.11M$ gates per second for non-XOR and XOR gates, respectively.

### 4.5 Empirical Analysis

Table 4 details DeepSecure performance in the realization of four different DL benchmarks without including the data and DL network pre-processing. Our benchmarks include both DNN and CNN models for analyzing visual, audio, and smart-sensing datasets. The topology of each benchmark is outlined in the following.

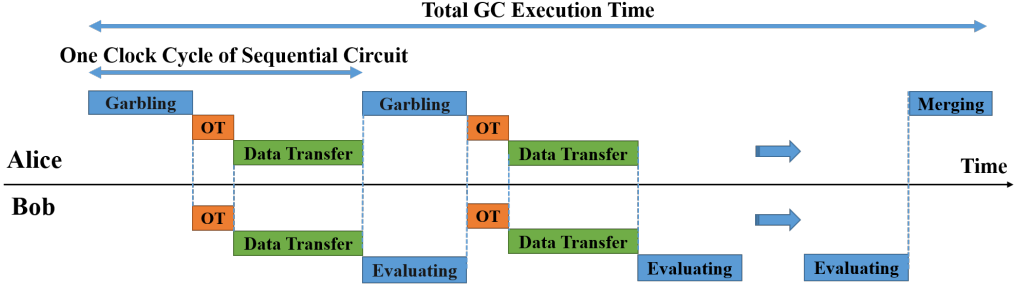

Figure 5: The timing diagram of GC execution in DeepSecure.

Table 4: Number of XOR and non-XOR gates, communication, computation time, and overall execution time for our benchmarks without involving the data and DL network pre-processing.

| Name | Network Architecture | #XOR | #Non-XOR | Comm. (MB) | Comp. (s) | Execution (s) |
| --- | --- | --- | --- | --- | --- | --- |
| Benchmark 1 | 28×28-5C2-ReLu-100FC-ReLu-10FC-Softmax | 4.31E7 | 2.47E7 | 7.91E2 | 1.98 | 9.67 |
| Benchmark 2 | 28×28-300FC-Sigmoid-100FC-Sigmoid-10FC-Softmax | 1.09E8 | 6.23E7 | 1.99E3 | 4.99 | 24.37 |
| Benchmark 3 | 617-50FC-Tanh-26FC-Softmax | 1.32E7 | 7.54E6 | 2.41E2 | 0.60 | 2.95 |
| Benchmark 4 | 5625-2000FC-Tanh-500FC-Tanh-19FC-Softmax | 4.89E9 | 2.81E9 | 8.98E4 | 224.50 | 1098.3 |

Table 5: Number of XOR and non-XOR gates, communication, computation time, and overall execution time for our benchmarks after considering the pre-processing steps. The last column of the table denotes the improvement achieved as a result of applying our pre-processing methodology.

| Name | Data and Network Compaction | #XOR | #Non-XOR | Comm. (MB) | Comp. (s) | Execution (s) | Improvement |
| --- | --- | --- | --- | --- | --- | --- | --- |
| Benchmark 1 | 9-fold | 4.81E6 | 2.76E6 | 8.82E1 | 0.22 | 1.08 | 8.95× |
| Benchmark 2 | 12-fold | 1.21E7 | 6.57E6 | 2.10E2 | 0.54 | 2.57 | 9.48× |
| Benchmark 3 | 6-fold | 2.51E6 | 1.40E6 | 4.47E1 | 0.11 | 0.56 | 5.27× |
| Benchmark 4 | 120-fold | 6.28E7 | 3.39E7 | 1.08E3 | 2.78 | 13.26 | 82.83× |

#### 4.5.1 Visual Benchmarks
Detecting objects in an image is a key enabler in devising various artificial intelligence and learning tasks. We evaluate DeepSecure practicability in analyzing MNIST dataset [33] using two different DL architectures. This data contains hand-written digits represented as $28 \times 28$ pixel grids, where each pixel is denoted by a gray level value in the range of 0-255.

**Benchmark 1.** In this experiment, we train and use a 5-layer convolutional neural network for document classification as suggested in [8]. The five layers include: (i) a convolutional layer with a kernel of size $5 \times 5$, a stride of (2, 2), and a map-count of 5. This layer outputs a matrix of size $5 \times 13 \times 13$. (ii) A ReLu layer as the non-linearity activation function. (iii) A fully-connected layer that maps the ($5 \times 13 \times 13 = 865$) units computed in the previous layers to a 100-dimensional vector. (iv) Another ReLu non-linearity layer, followed by (v) a final fully-connected layer of size 10 to compute the probability of each inference class.

**Benchmark 2.** In this experiment, we train and use LeNet-300-100 as described in [34]. LeNet-300-100 is a classical feed-forward neural network consisting of three fully-connected layers interleaved with two non-linearity layers (Sigmoid) with total 267K DL parameters.

#### 4.5.2 Audio Benchmark
**Benchmark 3.** Processing audio data is an important step in devising different voice activated learning tasks that appear in mobile sensing, robotics, and autonomous applications. Our audio data collection consists of approximately 1.25 hours of speech collected by 150 speakers [35]. In this experiment, we train and use a 3-layer fully-connected DNN of size ($617 \times 50 \times 26$) with Tanh as the non-linear activation function to analyze data within 5% inference error as suggested in [24].

#### 4.5.3 Smart-Sensing Benchmark
**Benchmark 4.** Analyzing smart-sensing data collected by embedded sensors such as accelerometers and gyroscopes is a common step in the realization of various learning tasks. In our smart-sensing data analysis, we train and use a 4-layer fully-connected DNN of size ($5625 \times 2000 \times 500 \times 19$) with Tanh as the non-linear activation function to classify 19 different activities [36] within 5% inference error as suggested in [24].

## 4.6 DeepSecure Pre-processing Effect

Table 5 shows DeepSecure performance for each benchmark after including the data and DL network pre-processing. Our pre-processing customization is an arbitrary step that can be used to minimize the number of required XOR and non-XOR gates for the realization of a particular DL model. As illustrated, our pre-processing approach reduces the execution time of GC protocol by up to 82-fold without any drop in the underlying DL accuracy.

## 4.7 Comparison with Prior Art Framework

Table 6 details the computation and communication overhead per sample in DeepSecure framework compared with the prior art privacy-preserving DL system [8]. Our result shows more than 58-fold improvement in terms of overall execution time per sample even without considering the pre-processing steps. For instance, it takes 570.11 seconds to run a single instance on the pertinent MNIST network using [8] while DeepSecure reduces this time to 9.67 seconds with no data and network pre-processing. Our data and DL network pre-processing further reduces the processing time per sample to only 1.08 seconds with no drop in the target accuracy.[3] As we discussed in Section 3.7, the confidentiality level of data samples and DL parameters does not change as a result of employing our pre-processing techniques.

**Table 6: Communication and computation overhead per sample in DeepSecure versus CryptoNet [8] for benchmark 1.**

| Framework | Comm. | Comp. (s) | Execution (s) | Improvement |
|---|---|---|---|---|
| DeepSecure without pre-processing | 791MB | 1.98 | 9.67 | 58.96 × |
| DeepSecure with pre-processing | 88.2MB | 0.22 | 1.08 | 527.88× |
| CryptoNets | 74KB | 570.11 | 570.11 | - |

Figure 6 shows the expected processing time as a function of data batch size from the client's point of view. The reported runtime for CryptoNet corresponds to implementing benchmark 1 using 5-10 bit precision on a Xeon E5-1620 CPU running at 3.5GHz, with 16GB of RAM as presented in [8]. Whereas, DeepSecure is prototyped using 16 bit number representation on an intel Core-i7 processor that has a slightly less computing power compared to the Xeon processor [37].

As illustrated in Figure 6, DeepSecure's computational cost scales linearly with respect to the number of samples. As such, DeepSecure is particularly ideal for scenarios in which *distributed clients stream small batches of data* (e.g., $N_{client} \leq 2590$) and send them to the server to find the corresponding inference label with *minimal delay*. However, CryptoNet is better-suited for settings where one client has a large batch of data (e.g., $N_{client} = 8192$) to process at once. This is because CryptoNet incurs a constant computational cost up to a certain number of samples depending on the choice of the polynomial degree. To mitigate the cost, authors in [8] suggest processing data in batches as opposed to individual samples using scalar encoding. The data batch size, in turn, is dictated by the polynomial degree used in the realization of a particular DL model. Therefore, to acquire a higher security level one might need to use larger data batch sizes in the CryptoNet framework.

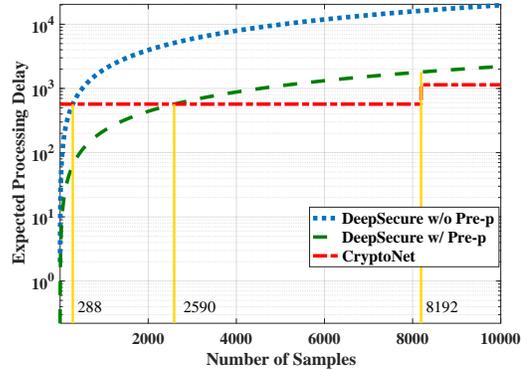

**Figure 6: Expected processing time from client's point of view as a function of data batch size. In this figure, the $y$ axis is illustrated in logarithmic scale.**

## 5 RELATED WORK

Authors in [38] have suggested the use of secure function evaluation protocols to securely evaluate a DL model. Their proposed approach, however, is an interactive protocol in which the data owner needs to first encrypt the data and send it to the cloud. Then, the cloud server should multiply the encrypted data with the weights of the first layer, and send back the results to the data owner. The data owner decrypts, applies the pertinent non-linearity, and encrypts the result again to send it back to the cloud server for the evaluation of the second layer of the DL model. This process continues until all the layers are computed. There are several limitations with this work [38]: (i) it leaks partial information embedded in the weights of the DL model to the data owner. (ii) It requires the data owner to have a constant connection with the cloud server while evaluating the DL network. To address the issue of information leakage as a result of sharing the intermediate results, [39] and [40] enhance the protocol initially proposed in [38] to obscure the weights. However, even these works [39, 40] still need to establish a constant connection with the client to delegate the non-linearity computations after each hidden layer to the data owner and do not provide the same level of security provided by DeepSecure.

Perhaps the closest work to DeepSecure is the study by Microsoft Research [8] in which homomorphic encryption is used as the primary tool for privacy-preserving computation of DL networks. Unlike DeepSecure, the inherent noise in HE yields a trade-off between privacy and accuracy in evaluating the DL model which in turn translates to a lower accuracy level for obtaining a higher degree of privacy. In addition, the relatively high computation overhead of HE bounds the applicability of such approach to the use of low-degree polynomials and limited-precision numbers (e.g., 5-10 bits). To the best of our knowledge, DeepSecure is the first to propose a scalable, fully-secure model for DL evaluation while limiting the communications between the client and the cloud

---

[3] Authors in [8] have only reported one benchmark (which we used as *benchmark 1*) for proof-of-concept evaluation. As such, we did not include the comparison for the other three benchmarks used in this paper.

server to a constant number regardless of the input data and DL network size.

A number of earlier works have shown the usability of data projection and sparsity regularization techniques to facilitate feature extraction [26, 27, 41] or accelerate the execution of particular DL models [24, 28]. These set of works have been mainly focused on the functionality of DL models in terms of the accuracy and physical performance (e.g., energy consumption, memory footprint, etc.) with no attention to the data privacy. To the best of our knowledge, DeepSecure is the first framework that introduces, implements, and automates the idea of data and DL network transformation as a way to minimize the number of required non-XOR gates for the privacy-preserving realization of DL models using GC.

## 6 CONCLUSION

We present DeepSecure, a novel practical and provably-secure DL framework that enables distributed clients (data owners) and cloud servers (who have the capability and resources to tune large scale DL models), jointly evaluate a DL network on their private assets. DeepSecure leverages automated design, efficient logic synthesis tools, and optimization methodologies to provide scalable realization of functions required for DL evaluation optimized for Yao's GC protocol. We also provide mechanisms to securely outsource the GC computation in settings where clients incur severe resource constraints. Our GC-optimized realization of hierarchical non-linear DL models demonstrates more than 58 times higher throughput per sample compared with the prior art privacy-preserving DL solution. We further propose a set of data and DL network transformation techniques as a pre-processing step to explicitly optimize the computation and communication overhead of GC protocol in the context of deep learning. Proof-of-concept evaluations using different DL benchmarks shows up to two orders-of-magnitude additional improvements achieved as a result of our pre-processing methodology.

## REFERENCES

[1] Li Deng and Dong Yu. Deep learning: methods and applications. *Foundations and Trends in Signal Processing*, 7(3–4), 2014.
[2] Yann LeCun, Yoshua Bengio, and Geoffrey Hinton. Deep learning. *Nature*, 521(7553), 2015.
[3] Nicola Jones et al. The learning machines. *Nature*, 505(7482):146–148, 2014.
[4] Jeremy Kirk. Ibm join forces to build a brain-like computer. http://www.pcworld.com/article/2051501/universities-join-ibm-in-cognitive-computing-researchproject.html, 2016.
[5] Amir Efrati. How "deep learning" works at apple, beyond. https://www.theinformation.com/How-Deep-Learning-Works-at-Apple-Beyond, 2017.
[6] Reza Shokri and Vitaly Shmatikov. Privacy-preserving deep learning. In *Proceedings of the 22nd ACM SIGSAC Conference on Computer and Communications Security*, pages 1310–1321. ACM, 2015.
[7] Martín Abadi, Andy Chu, Ian Goodfellow, H Brendan McMahan, Ilya Mironov, Kunal Talwar, and Li Zhang. Deep learning with differential privacy. *arXiv preprint arXiv:1607.00133*, 2016.
[8] Ran Gilad-Bachrach, Nathan Dowlin, Kim Laine, Kristin Lauter, Michael Naehrig, and John Wernsing. Cryptonets: Applying neural networks to encrypted data with high throughput and accuracy. In *Proceedings of The 33rd International Conference on Machine Learning*, pages 201–210, 2016.
[9] Ebrahim M Songhori, Siam U Hussain, Ahmad-Reza Sadeghi, Thomas Schneider, and Farinaz Koushanfar. Tinygarble: Highly compressed and scalable sequential garbled circuits. In *2015 IEEE Symposium on Security and Privacy*, pages 411–428. IEEE, 2015.
[10] Alex Krizhevsky, Ilya Sutskever, and Geoffrey E Hinton. Imagenet classification with deep convolutional neural networks. In *Advances in neural information processing systems*, pages 1097–1105, 2012.
[11] Moni Naor and Benny Pinkas. Computationally secure oblivious transfer. *Journal of Cryptology*, 18(1):1–35, 2005.
[12] Andrew Chi-Chih Yao. How to generate and exchange secrets. In *Foundations of Computer Science, 1986., 27th Annual Symposium on*, pages 162–167. IEEE, 1986.
[13] Mihir Bellare, Viet Tung Hoang, Sriram Keelveedhi, and Phillip Rogaway. Efficient garbling from a fixed-key blockcipher. In *Security and Privacy (SP), 2013 IEEE Symposium on*, pages 478–492. IEEE, 2013.
[14] Moni Naor, Benny Pinkas, and Reuban Sumner. Privacy preserving auctions and mechanism design. In *Proceedings of the 1st ACM conference on Electronic commerce*, pages 129–139. ACM, 1999.
[15] Vladimir Kolesnikov and Thomas Schneider. Improved garbled circuit: Free xor gates and applications. In *International Colloquium on Automata, Languages, and Programming*, pages 486–498. Springer, 2008.
[16] Samee Zahur, Mike Rosulek, and David Evans. Two halves make a whole. In *Annual International Conference on the Theory and Applications of Cryptographic Techniques*, pages 220–250. Springer, 2015.
[17] Ben Kreuter, Abhi Shelat, Benjamin Mood, and Kevin Butler. Pcf: A portable circuit format for scalable two-party secure computation. In *Presented as part of the 22nd USENIX Security Symposium (USENIX Security 13)*, pages 321–336, 2013.
[18] Yehuda Lindell and Benny Pinkas. An efficient protocol for secure two-party computation in the presence of malicious adversaries. In *EUROCRYPT'07*, volume 4515, pages 52–78. Springer, 2007.
[19] Jesper Buus Nielsen and Claudio Orlandi. LEGO for two-party secure computation. In *TCC'09*, volume 5444, pages 368–386. Springer, 2009.
[20] Abhi Shelat and Chih-hao Shen. Two-output secure computation with malicious adversaries. In *EUROCRYPT'11*, volume 6632, pages 386–405. Springer, 2011.
[21] Yehuda Lindell and Benny Pinkas. Secure two-party computation via cut-and-choose oblivious transfer. *Journal of cryptology*, 25(4):680–722, 2012.
[22] Eva L Dyer, Aswin C Sankaranarayanan, and Richard G Baraniuk. Greedy feature selection for subspace clustering. *Journal of Machine Learning Research*, 14(1):2487–2517, 2013.
[23] Azalia Mirhoseini, Bita Darvish Rouhani, Ebrahim M Songhori, and Farinaz Koushanfar. Perform-ml: Performance optimized machine learning by platform and content aware customization. In *Proceedings of the 53rd Annual Design Automation Conference*, page 20. ACM, 2016.
[24] Bita Darvish Rouhani, Azalia Mirhoseini, and Farinaz Koushanfar. Delight: Adding energy dimension to deep neural networks. In *Proceedings of the 2016 International Symposium on Low Power Electronics and Design*, pages 112–117. ACM, 2016.
[25] Bita Darvish Rouhani, Azalia Mirhoseini, and Farinaz Koushanfar. Deep3: Leveraging three levels of parallelism for efficient deep learning. In *Proceedings of the 54rd Annual Design Automation Conference*. ACM, 2017.
[26] Thomas Serre, Gabriel Kreiman, Minjoon Kouh, Charles Cadieu, Ulf Knoblich, and Tomaso Poggio. A quantitative theory of immediate visual recognition. *Progress in brain research*, 165:33–56, 2007.
[27] Wei Wen, Chunpeng Wu, Yandan Wang, Yiran Chen, and Hai Li. Learning structured sparsity in deep neural networks. In *Advances in Neural Information Processing Systems*, pages 2074–2082, 2016.
[28] Song Han, Jeff Pool, John Tran, and William Dally. Learning both weights and connections for efficient neural network. In *Advances in Neural Information Processing Systems*, pages 1135–1143, 2015.
[29] Mihir Bellare, Viet Tung Hoang, and Phillip Rogaway. Foundations of garbled circuits. In *Proceedings of the 2012 ACM conference on Computer and communications security*, pages 784–796. ACM, 2012.
[30] Christof Paar and Jan Pelzl. *Understanding cryptography: a textbook for students and practitioners*. Springer Science & Business Media, 2009.
[31] Synopsys inc. Design compiler. http://www.synopsys.com/Tools/Implementation/RTLSynthesis/DesignCompiler, 2000.
[32] Hesham Amin, K Memy Curtis, and Barrie R Hayes-Gill. Piecewise linear approximation applied to nonlinear function of a neural network. *IEE Proceedings-Circuits, Devices and Systems*, 144(6):313–317, 1997.
[33] Yann LeCun, Corinna Cortes, and Christopher Burges. Mnist dataset. http://yann.lecun.com/exdb/mnist/, 2017.
[34] Yann LeCun, Léon Bottou, Yoshua Bengio, and Patrick Haffner. Gradient-based learning applied to document recognition. *Proceedings of the IEEE*, 86(11):2278–2324, 1998.
[35] UCI machine learning repository. https://archive.ics.uci.edu/ml/datasets/isolet, 2017.
[36] UCI machine learning repository. https://archive.ics.uci.edu/ml/datasets/Daily+and+Sports+Activities, 2017.
[37] Intel Processors. http://www.velocitymicro.com/blog/xeon-vs-i7i5-whats-difference/, 2017.
[38] Mauro Barni, Claudio Orlandi, and Alessandro Piva. A privacy-preserving protocol for neural-network-based computation. In *Proceedings of the 8th workshop on Multimedia and security*, pages 146–151. ACM, 2006.

[39] Claudio Orlandi, Alessandro Piva, and Mauro Barni. Oblivious neural network computing via homomorphic encryption. *EURASIP Journal on Information Security*, 2007:18, 2007.

[40] Alessandro Piva, Claudio Orlandi, M Caini, Tiziano Bianchi, and Mauro Barni. Enhancing privacy in remote data classification. In *IFIP International Information Security Conference*, pages 33–46. Springer, 2008.

[41] Azalia Mirhoseini, Eva L Dyer, Ebrahim Songhori, Richard G Baraniuk, Farinaz Koushanfar, et al. Rankmap: A platform-aware framework for distributed learning from dense datasets. *arXiv preprint arXiv:1503.08169*, 2015.